\input{aipcheck}

\documentclass[final]{aipproc}

\layoutstyle{6x9}

\RequirePackage{xspace}
 
\usepackage{relsize}
\def\babar{\mbox{\slshape B\kern-0.1em{\smaller A}\kern-0.1em
    B\kern-0.1em{\smaller A\kern-0.2em R}}}
\def\ccbar {\ensuremath{c\overline c}\xspace}
\def\pep2{PEP-II}
\def\BF{$B$ Factory}

\def\Y#1S{\ensuremath{\Upsilon{(#1S)}}\xspace}% no space before {...}!
\def\FourS {\Y4S}
\newcommand{\mev}{\ensuremath{\mathrm{\,Me\kern -0.1em V}}\xspace}
\newcommand{\gevc}{\ensuremath{{\mathrm{\,Ge\kern -0.1em V\!/}c}}\xspace}
\newcommand{\mevc}{\ensuremath{{\mathrm{\,Me\kern -0.1em V\!/}c}}\xspace}
\newcommand{\gevcc}{\ensuremath{{\mathrm{\,Ge\kern -0.1em V\!/}c^2}}\xspace}
\newcommand{\mevcc}{\ensuremath{{\mathrm{\,Me\kern -0.1em V\!/}c^2}}\xspace}
\def\to                 {\ensuremath{\rightarrow}\xspace}
\newcommand{\stat}{\ensuremath{\mathrm{(stat)}}\xspace}
\newcommand{\syst}{\ensuremath{\mathrm{(syst)}}\xspace}
\def\Dbar    {\kern 0.2em\overline{\kern -0.2em D}{}\xspace}

\def\Dz      {\ensuremath{D^0}\xspace}
\def\Dzb     {\ensuremath{\Dbar^0}\xspace}

\def\Dstarp  {\ensuremath{D^{*+}}\xspace}

\begin{document}
\title{Charm Physics at $\babar$}


\author{Chunhui Chen}
{address={Department of Physics, University of Maryland\\
College Park, Maryland 20742-4111, U.S.A\\
(for the \babar\ Collaboration)}}

\begin{abstract}
Large production of the $\ccbar$ pairs and high integrated luminosity
make the \pep2 \BF\, an excellent place for studying the charm hadrons.
In this paper, we present a few of the most recent results from 
the $\babar$ collaboration in the charm sector.
\end{abstract}

\maketitle

\section{Introduction}
The $\babar$ detector~\cite{Aubert:2001tu} is a general 
purpose detector designed to collect
data at \pep2 asymmetric $e^+e^-$ collider, operating at the
center-of-mass energy corresponding to the \FourS resonance or 
$\sim 40\,\mev$ below it.
With copious production of $\ccbar$ pairs from the continuum and
high integrated luminosity, $\babar$ is not only a \BF, it is also an
excellent laboratory to study the charm production and decays. In this 
paper, we present a few of the most recent charm analysis results from 
$\babar$.

\section{$\mathbf{\Dz-\Dzb}$ mixing }
Charm mixing is characterized by two dimensionless parameters, 
$x\equiv \Delta m/\Gamma$ and $y\equiv \Delta\Gamma/2\Gamma$, 
where $\Delta m$ ($\Delta\Gamma$) is the mass (width) difference between 
the two neutral $D$ mass eigenstates, and $\Gamma$ is the average 
width. If either $x$ or $y$ is nonzero, then the 
$\Dz-\Dzb$ mixing will occur. In the Standard Model (SM), $\Dz-\Dzb$ mixing
rate is heavily suppressed by the Glashow-Iliopoulos-Maiani 
(GIM) mechanism\cite{Glashow:gm}.
However, the SM mixing rate can be enhanced by the non-perturbative
effects and possible new physics beyond SM.

Based on a sample of $87\,\mbox{fb}^{-1}$ data, $\babar$ performed
a search of $\Dz-\Dzb$ mixing~\cite{Aubert:2004bn} to measure the 
overall time-integrated mixing rate $R_{mix}=(x^2+y^2)/2$ 
using the decay chain
$D^{*+}\to D^0\pi^+,\;D^0\to K^{\pm}e^\mp\nu$~\cite{conjugate}. 
The charge of the pion
daughter of the charged $D^*$ identifies the production flavor
of the neutral $D$, while the charge of the electron identifies the
decay flavor. These charges are equal for unmixed decays and
opposite for mixed decay. 
Using event selection and reconstruction based on neural networks
and charged kaon and electron 
particle identification, we obtained $49620\pm265$ unmixed events
and $114\pm 61$ mixed events. This results in
\begin{equation}
\begin{array}{lcl}
R_{mix}&=& 0.0024\pm 0.0012\stat\pm 0.0004 \syst\\
R_{mix}&<& 0.0042\;\mbox{at}\;90\,\%\;\mbox{CL.}
\end{array}
\end{equation}

\section{Search for $D^0\to\ell^+\ell^-$ }
In the SM, the flavor-changing neutral current (FCNC)
decays $\Dz\to e^+e^-$ and $\Dz\to\mu^+\mu^-$
are highly suppressed by the GIM mechanism.
The lepton-flavor violating (LFV) decay $\Dz\to e^\pm\mu^\mp$ is
strictly forbidden in the SM.
Some extensions to the Standard Model~\cite{Burdman:2001tf}, 
such as the $R$-parity violating supersymmetry, can enhance 
the FCNC processes by many orders of magnitude and can also permit
the LFV decays. 

BaBar performed a search for the decays of 
$\Dz\to e^+e^-$, $\Dz\to\mu^+\mu^-$, and
$\Dz\to e^\pm\mu^\mp$ based on a sample of 
$122\,\mbox{fb}^{-1}$  data~\cite{Aubert:2004bs}. 
To ensure as clean a sample 
as possible, the reconstructed $D^0\to\ell^+\ell^-$
candidate is required to originate from a $\Dstarp\to\Dz\pi^+$ decay.
A minimum value of 2.4\gevc is imposed on the
center-of-mass momentum of each \Dz candidate to further
reduce the combinatorial background
involving the decay products of $B$ mesons.
Tight particle selection criteria are also applied
to the daughters of $D^0\to\ell^+\ell^-$ decays.

We observed no significant signals in all three decay modes. As
a result, the branching fraction upper limits (UL) have been 
calculated using the $D^0\to\pi^+\pi^-$ decay as the
normalization mode. We obtain
\begin{equation}
\begin{array}{lcl}
\mbox{Br}(D^0\to e^+e^-) &<& 1.2\times 10^{-6}
\;\mbox{at}\;90\,\%\;\mbox{CL.},\\
\mbox{Br}(D^0\to \mu^+\mu^-) &<& 1.3\times 10^{-6}
\;\mbox{at}\;90\,\%\;\mbox{CL.},\\
\mbox{Br}(D^0\to \mu^\pm e^\mp) &<& 8.1\times 10^{-7}
\;\mbox{at}\;90\,\%\;\mbox{CL.}.
\end{array}
\end{equation}
These results represent significant improvements on the previous 
limits~\cite{Aitala:1999db,Abt:2004hn}.

\section{Search for $D_{sJ}(2632)^+$}
\begin{figure}[htb]
\includegraphics[height=5.2cm,width=5.2cm]{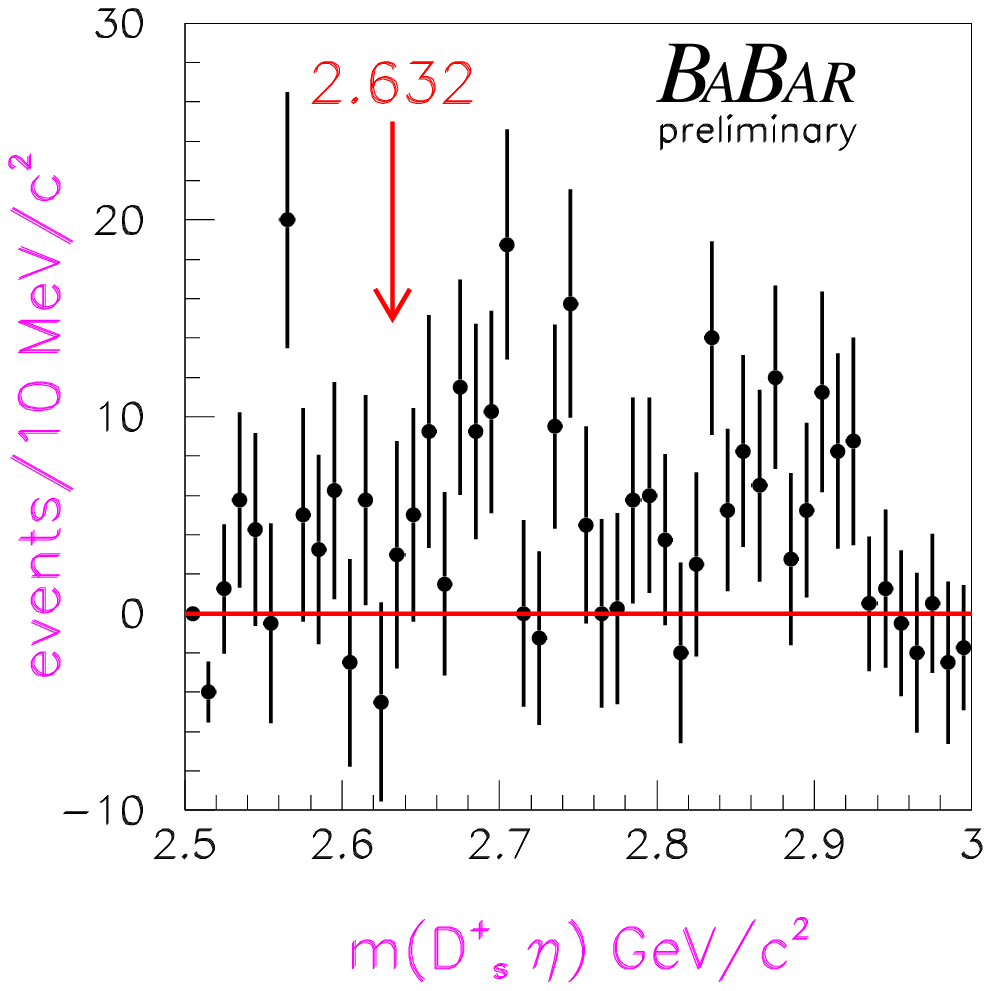}
\includegraphics[height=4.5cm,width=4.5cm]{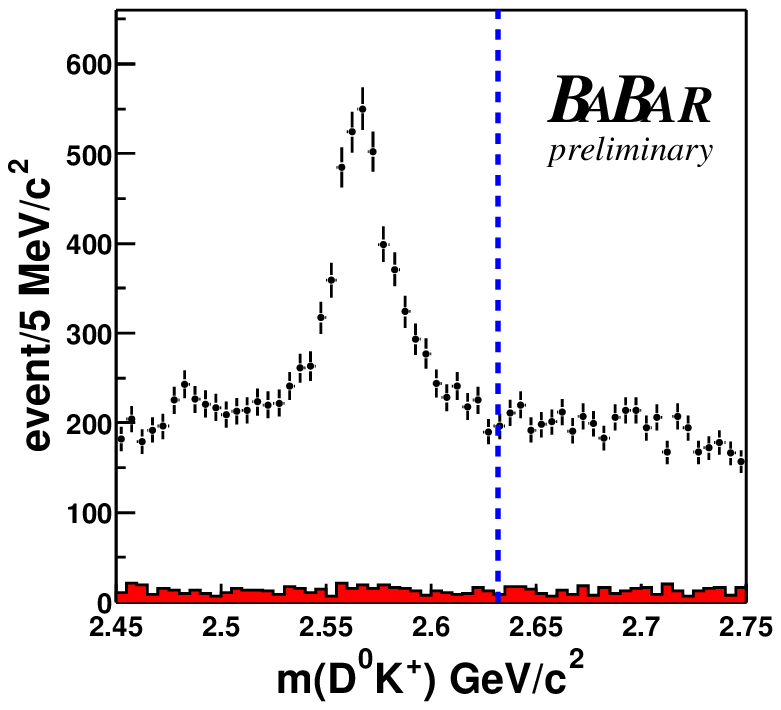}
\includegraphics[height=4.5cm,width=4.5cm]{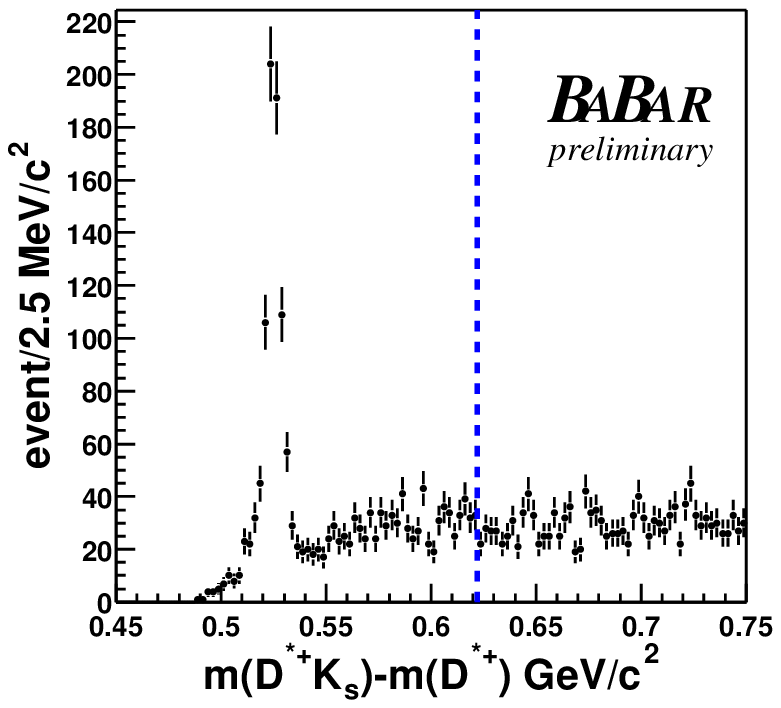}
\caption{\label{fig:DsJ2632}
(Left:) The $D^+_s\eta$ invariant mass distribution after 
background subtraction. The arrow indicates the mass location
of the expected $D_{sJ}(2632)^+$ state. (Middle:) The
$D^0K^+$ invariant mass distribution. The red histogram is
the invariant mass distribution of $D^0K^-$ combinations, and
the blue line indicates the mass location of the expected $D_{sJ}(2632)^+$ 
state. (Right:) The distribution of the difference in invariant mass 
of the $D^{*+}K^0_S$ combination and $D^{*+}$ candidate. 
The blue line indicates the mass location of the expected $D_{sJ}(2632)^+$ 
state.}
\end{figure}

The SELEX Collaboration at FNAL has recently reported the
observation of a narrow state~\cite{Evdokimov:2004iy} 
at a mass of 2632\,\mevcc that 
decays to $D^+_s\eta$. Evidence for the same state in the
 $D^0K^+$ mass spectrum was also presented.
BaBar has searched for this resonance in the final states
$D^+_s\eta$, $D^0K^+$ and $D^{*+}K^0_S$~\cite{Aubert:2004ku} 
produced in $e^+e^-\to\ccbar$ events, using $125\,\mbox{fb}^{-1}$ data.
As shown in Fig.~\ref{fig:DsJ2632},
no signal is observed in  $D^+_s\eta$ decay channel;
similarly, no evidence of $D_{sJ}(2632)^+$ was found in the
$D^0K^+$ and $D^{*+}K^0_S$ final states, although large and 
clean signals for the decay 
$D_{s2}(2573)^+\to D^0K^+$ and 
$D_{s1}(2536)^+\to D^{*+}K^0_S$ are seen.

\section{$\Xi^0_c$ Production and Decays}
Using a sample of $116\,\mbox{fb}^{-1}$ data, $\babar$ has
performed a branching fraction ratio measurement of the $\Xi^0_c$ 
decaying to $\Omega^-K^+$ and $\Xi^-\pi^+$~\cite{Aubert:2005cu}. 
The result
\begin{equation}
\frac{\mbox{Br}(\Xi^0_c\to\Omega^- K^+)}
{\mbox{Br}(\Xi^0_c\to\Xi^-\pi^+)}=0.294\pm0.018\,
\stat\pm 0.016\,\syst
\end{equation}
is a significant improvement over the previous measurement
by CLEO~\cite{Henderson:1992cx} and is consistent with a 
spectator quark model prediction~\cite{Korner:1992wi}.

Although copious production of $\Xi^0_c$ in $B$ decays has been
predicted, such process has been only observed by 
CLEO~\cite{Barish:1997pq} with a 
significance of $\sim 3\sigma$ in the $\Xi^0_c\to\Xi^-\pi^+$
decay mode and $\sim 4\sigma$ in the $\Xi^+_c\to\Xi^-\pi^+\pi^+$
decay mode. We studied the $\Xi^0_c$ production by
measuring the spectrum of the $\Xi^0_c$ momentum $p^*$ in the
$e^+e^-$ center-of-mass frame. The $\Xi^0_c$ produced by the
$B$ decays tends to have a smaller momentum; its $p^*$ distribution
peaks below $1.5\,\gevc$ and has a kinematic limit of 
$p^{*}=2.135\,\gevc$  at $\babar$. As for the $\Xi^0_c$ from continuum 
production, its momentum distribution peaks at a much higher
$p^{*}$ value. By examining the $p^{*}$ distribution of the
$\Xi^0_c$ from  on-resonance and off-resonance sample, we
found
\begin{equation}
\mbox{Br}(B\to\Xi^0_c X)\times 
\mbox{Br}(\Xi^0_c\to\Xi^-\pi^+)=2.11\pm0.19\,\stat
\pm0.25\,\syst\times 10^{-4},
\end{equation}
and
\begin{equation}
\sigma(e^+e^-\to\Xi^0_cX)\times
\mbox{Br}(\Xi^0_c\to\Sigma^-\pi^+)=
388\pm39\,\stat\pm41\,\syst\,\mbox{fb},
\end{equation}
where both $\Xi^0_c$ and $\bar{\Xi}^0_c$ are included
in the cross-section.

\section{Measurement of $\Lambda^+_c$ mass}
The invariant masses of the charm hadron ground states are currently
reported by the Particle Data Group (PDG) with a precision of about
0.5--0.6\mevcc~\cite{Eidelman:2004wy}. The best individual 
measurements have a
statistical and systematic precision of about $0.5$\mevcc and use data
samples of a few hundred events.  The
\babar~data sample contains a large amount of different charm hadron
decays and, due to the excellent momentum and vertex resolution in
\babar, many of the decay modes can be reconstructed with an
event-by-event mass uncertainty of a few \mevcc. We can therefore
provide significantly improved estimate of the charm hadron masses.

With a sample of $232\,\mbox{fb}^{-1}$ data, $\babar$ 
performed a precision measurement of the $\Lambda^+_c$
mass. The measurement is based on the reconstruction of the decay
modes $\Lambda^+_c\to\Lambda\bar{K}^0K^+$ and 
$\Lambda^+_c\to\Sigma^0\bar{K}^0K^+$. Because almost all of the
$\Lambda^+_c$ invariant mass in these decays tied to the well-known 
rest masses of the $\Lambda^+_c$ decay products, the systematic 
uncertainty in
the reconstructed mass is significantly reduced compared to the
measurements that try to use the more common decay modes.
Combining the results 
from those two modes, the measured $\Lambda^+_c$ mass
is
\begin{equation}
m(\Lambda^+_c)=2286.46\pm0.14\,\mevcc.
\end{equation}
This result is in agreement with the mass values measured in other 
much large sample of $\Lambda^+_c$ decays, including
$\Lambda^+_c\to pK^-\pi^+$ and $\Lambda^+_c\to pK^0_S$ decays, 
although these are subject to large systematic uncertainties. 

This $\Lambda^+_c$ mass measurement is the most precise measurement 
of an open charm hadron mass to date and is an improvement in precision 
by more than a factor of four over the current PDG value of
$2284.9\pm 0.6\mevcc$. Our result is about $2.5\sigma$ higher
than the PDG value, which is based on several high $Q$-value
decay modes, mainly $\Lambda^+_c\to pK^-\pi^+$ decays.

\section{Conclusion}
$\babar$ has a very rich and active charm physics program.
In this paper we discussed only a few most recent results
from $\babar$. Given the excellent luminosity achieved 
by \pep2, much more high precision charm physics results 
are expected in the near future.

\bibliographystyle{aipproc}

\begin{thebibliography}{9}
\bibitem{conjugate}
We imply charged conjugate modes throughout the paper.
\end{thebibliography}

\end{document}